\documentclass[pra,twocolumn,showpacs,nobibnotes,floatfix,superscriptaddress]{revtex4}

\usepackage{latexsym,amsmath,amssymb,amsfonts,mathbbol,graphicx,color}
\usepackage{dcolumn}
\usepackage{bm}

\begin{document}

\title{Entanglement Sudden Death as an Indicator of Fidelity in a Four-Qubit Cluster State}

\author{Yaakov S. Weinstein}
\affiliation{Quantum Information Science Group, {\sc Mitre},
260 Industrial Way West, Eatontown, NJ 07224, USA}


\begin{abstract}
I explore the entanglement evolution of a four qubit cluster state in a dephasing 
environment concentrating on the phenomenon of entanglement sudden death (ESD).
Specifically, I ask whether the onset of ESD has an effect on the utilization of this 
cluster state as a means of implementing a single qubit rotation in the measurement based 
cluster state model of quantum computation. To do this I compare 
the evolution of the entanglement to the fidelity, a measure of how accurately  
the desired state (after the measurement based operations) is achieved. I find that 
ESD does not cause a change of behavior or discontinuity in the fidelity but may indicate
when the fidelity of certain states goes to .5.
\end{abstract}


\maketitle

\section{Introduction}

Entanglement is a uniquely quantum mechanical phenomenon in which quantum 
systems exhibit correlations above and beyond what is possible for classical systems. 
Entangled systems are thus an important resource for many quantum information processing 
protocols including quantum computation, quantum metrology, and quantum communication 
\cite{book}. Much work has been done with respect to the identification and 
quantification of entanglement as well as explorations of entanglement evolution 
under a range of possible dynamics \cite{HHH}.

An important area of research is to understand the possible degredation of 
entanglement under decoherence. Decoherence, unwanted interactions between the system
and environment, is the major challenge confronting experimental 
implementations of quantum computation, metrology, and communication. 
Decoherence may be especially detrimental to highly non-classical,
and hence the most potentially useful, entangled states \cite{Dur}. A manifestation 
of the detrimental affects of decoherence on entangled states is entanglement sudden 
death (ESD) in which entanglement is completely lost in a finite time \cite{DH,YE1} 
despite the fact that the loss of system coherence 
is asymptotic. This aspect of entanglement has been well explored in the case of 
bi-partite systems and there are a number of studies looking at ESD in multi-partite 
systems \cite{SB,ACCAD,LRLSR,YYE,QECESD,YSW}. In addition, there have been several initial 
experimental ESD studies \cite{expt}. 

The ESD phenomenon is interesting on a fundamental level and important for the general 
study of entanglement. However, it is not yet clear what the affect of ESD is
on quantum information protocols. Are different quantum protocols helped, hurt, 
or indifferent to ESD? Previous studies along these lines have been in the area
of quantum error correction (QEC). An explicit study of the three-qubit phase
flip code concludes that this specific code is indifferent to ESD \cite{YSW}. In this paper I 
take a first step in studying the affect of ESD on cluster state quantum computational gates.
Specifically, I study a four qubit cluster state to see how ESD affects its utility
as a means of implementing a general single qubit rotation for measurement based (cluster state)
quantum computation. My approach will be to use an entanglement witness, the negativity
and bi-partite concurrence as entanglement metrics and compare the behavior of these 
metrics under the influence of decoherence to the fidelity of the final state after the 
attempted single qubit rotation. In addition, I will study the entanglement that remains in 
the cluster state after two measurements and compare it to the fidelity of 
the state of the two unmeasured qubits. 

The cluster state \cite{BR3} is a specific type of entangled state that can be used 
as an initial resource for a measurement based approach to quantum computation \cite{BR1}. 
A cluster state can be created by first rotating all qubits into the state 
$\frac{1}{\sqrt{2}}(|0\rangle + |1\rangle)$. Desired pairs of qubits are entangled by applying 
control phase (CZ) gates between them. In a graphical picture of a cluster state, qubits 
are represented by circles and pairs of qubits that have
been entangled via a CZ gate are connected by a line.  A cluster state
with qubits arranged in a two-dimensional lattice, such that each qubit
has been entangled with four nearest neighbors, suffices for universal QC.

After constructing the cluster state, any quantum computational algorithm
can be implemented using only single-qubit measurements along axes in the $x$-$y$
plane. These processing measurements are performed by column, from left 
to right, until only the last column is left unmeasured. The last column 
contains the output state of the quantum algorithm which can be extracted 
by a final readout measurement. One can view each row of the 
cluster-state lattice as the evolution of a single logical qubit in time.
Two (logical) qubit gates are performed via a connection between two rows of 
the cluster state. CZ gates in particular are `built-in' to the cluster state
and simple measurement automatically implements the gate. 
Single qubit rotations can be performed when there is no conncetion between 
the measured qubit(s) and qubits in another row. In such a case the logical 
gate implemented by measurement along an angle $\phi$ in the $x$-$y$ plane 
is $X(\pi m)HZ(\phi)$, where $H$ is the Hadamard gate and $Z(\alpha)$ ($X(\alpha$)) 
is a $z$- ($x$-) rotation by an angle $\alpha$ \cite{BR2}. The dependence of the
logical operation on the outcome of the measurement is manifest in $m = 0, 1$ 
for measurement outcome $-1, +1$. An arbitrary single qubit rotation can
be implemented via three logical single-qubit rotations of the above
sort yielding
\begin{equation}
HZ(\alpha+\pi m_{\alpha})X(\beta + \pi m_{\beta})
Z(\gamma + \pi m_{\gamma}), \nonumber
\end{equation}
where $(\alpha, \beta, \gamma)$ are the Euler angles of the
rotation. For example, by drawing the Euler angles according to the Haar measure, a
random single-qubit rotation can be implemented.
 
As with all quantum computing paradigms, cluster state quantum computation, 
both during the constuction of the cluster state and during subsequent 
measurement, are subject to 
decoherence. We study a four qubit cluster chain, with no interaction 
between the qubits (beyond the initial conditional phase gates used
to construct the cluster state) placed in a dephasing environment 
fully described by the Kraus operators
\begin{equation}
K_1 = \left(
\begin{array}{cc}
1 & 0 \\
0 & \sqrt{1-p} \\
\end{array}
\right); \;\;\;\;
K_2 = \left(
\begin{array}{cc}
0 & 0 \\
0 & \sqrt{p} \\
\end{array}
\right)
\end{equation} 
where we have defined the dephasing parameter $p$. When all four 
qubits undergo dephasing we have 16 Kraus operators each of the form 
$A_l = (K_i\otimes K_j\otimes K_k\otimes K_{\ell})$ where 
$l = 1,2,...,16$ and $i,j,k,\ell = 1,2$. Though all of the below calculations
are done with respect to $p$, I implicitly assume that $p$
increases with time, $\tau$, at a rate $\kappa$, such that 
$p = 1-e^{-\kappa\tau}$ and $p\rightarrow 1$ only at infinite times. 
For now I also assume equal dephasing for all four qubits. 

In optical cluster state construction small (few qubit) cluster states are fused together to
form larger cluster states \cite{BR}. The smaller states must be stored until they are needed
and may be subject to decohence (especially dephasing). In other cluster state implementations,
where complete two-dimensional cluster states can be constructed in just a few steps \cite{ExpC}, 
any four qubit chain may be attached to at least one other qubit. In this case 
our results may not be exact. 

While entanglement is invariant to single qubit operations, decoherence is not and 
local operations may play a significant role in the entanglement dynamics of the state. 
Thus, if a cluster state must be stored in a decohering environment one would ideally like 
to choose a cluster state representation (within single qubit operations) that has the greatest 
immunity to the decoherence so as retain as much entanglement 
as possible. With this is mind a secondary aim of this paper is to study two representations
of the four qubit chain cluster state and compare the affects of dephasing on these 
representations. The first representation of the four qubit cluster state is 
\begin{equation}
|C_4\rangle = \frac{1}{2}\left(|0000\rangle+|0011\rangle+|1100\rangle-|1111\rangle\right).
\end{equation}
This representation minimizes the number of computational basis states having non-zero 
contribution. The second representation is:
\begin{equation}
|C_{4H}\rangle = H_1H_4|C_4\rangle,
\end{equation}
where $H_j$ is the single qubit Hadamard gate on qubit $j$. This is the state one 
would get by initially rotating each qubit into the state $\frac{1}{\sqrt{2}}(|0\rangle+|1\rangle)$
and applying contolled phase gates $CZ_{12}$, $CZ_{23}$, and $CZ_{34}$. We note that `connections' 
between qubits may be added or removed by single qubit rotations (though the entanglement
stays constant) thus changing the operation performed via measurement \cite{box}.  

The four qubit cluster has pure four qubit entanglement.
Thus, for example, there is no bi-partite concurrence between any of the two
qubits. As an entanglement metric we use the negativity, $N$, for which we will 
simply use the most negative eigenvalue of the parital transpose of the density 
matrix \cite{neg}. There are a number of inequivalent forms of the negativity for the four
qubit cluster state: the partial transpose may be taken with respect to any single 
qubit, $N_1$, or the partial transpose may be taken with respect to two qubits: 
qubits 1 and 2, $N_{12}$, qubits 1 and 3, $N_{13}$, or qubits 1 and 4, $N_{14}$. 

A further method of monitoring entanglement evolution is via 
the expectation value of the state with respect to an appropriate 
entanglement witness \cite{EW}. Entanglement witnesses are observables with 
positive or zero expectation value for all states not in a specified 
class and a negative expectation value for at least one state of 
the specified class. Entanglement 
witnesses may allow for an efficient means of determining whether entanglement 
is present in a state (as opposed to inefficient state tomography). This is 
especially important for experimental implementations as it may be 
the only practical means of deciding whether or not sufficient entanglement 
is present in the system.  The entanglement witnesses I use are designed 
to detect cluster states and will be either 
$\mathcal{W}_{C4} = \openone/2-|C_4\rangle\langle C_4|$ or 
$\mathcal{W}_{C4H} = \openone/2-|C_{4H}\rangle\langle C_{4H}|$ depending
on the representation \cite{TG}. 

\section{ESD in a Four Qubit Cluster State}

\begin{figure}[t]
\includegraphics[width=4cm]{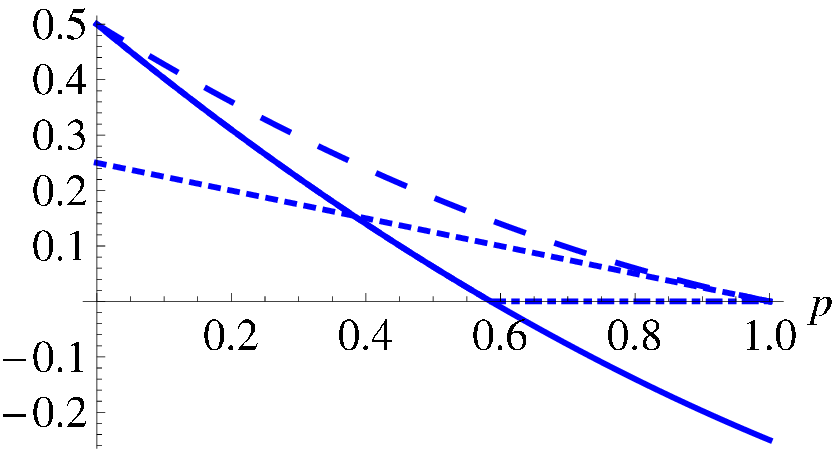}
\includegraphics[width=4cm]{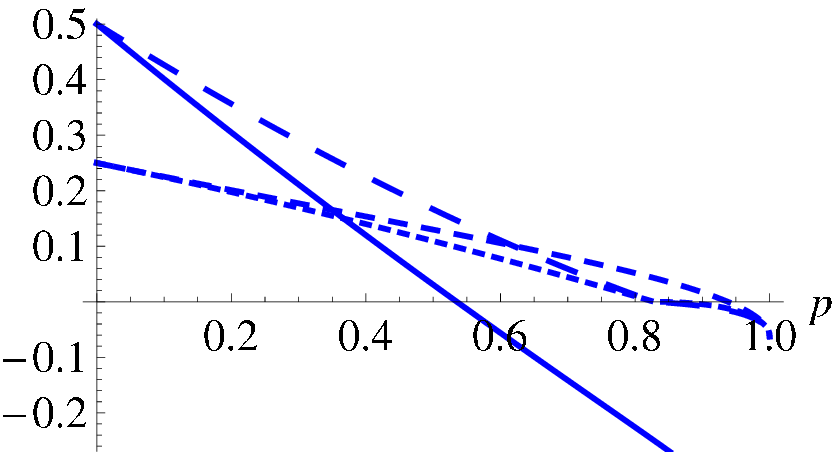}
\caption{\label{C4} (Color online)
Entanglement evolution as measured by $-\rm{Tr}[\mathcal{W}_r\rho_r(p)]$
(solid line), $N_1$ (large dashed line), $N_{12}$ (chained line),
$N_{13}$ (medium dashed line), and $N_{14}$ (small dashed line)  for 
intial states $|C_4\rangle$ (left) and $|C_{4H}\rangle$ (right) as a
function of dephasing strength $p$ on all four qubits. For intial state
$|C_{4}\rangle$ there is no ESD for $N_1$ or $N_{13} = N_{14}$, but ESD 
is exhibited for $N_{12}$ at $p \simeq .586$. The expectation value of the 
dephased state with respect to the entanglement witness $\mathcal{W}_{C4}$ 
is equivalent to $N_{12}$. 
For initial state 
$|C_{4H}\rangle$, $N_1 = N_{12}$ and ESD occurs at $p\simeq .828$. This 
is the same value for which $N_{14}$ exhibits ESD. ESD for $N_{13}$
is exhibited at $p = .938$. The entanglement witness, $\mathcal{W}_{C4H}$ 
fails to detect entanglement for $p \agt .535$.
}
\end{figure}
  
Our first step is to determine at what dephasing strength, $p$, (if any) the 
four qubit cluster state exhibits ESD. The final state of the four qubit system 
after dephasing is given by 
$\rho_{r}(p) = \sum_l^{16}A_l|C_r\rangle\langle C_r|A_l^{\dag}$ where $r = 4,4H$. 
Figure \ref{C4} shows the evolution of our chosen entanglement metrics for initial 
cluster states as a function of $p$. For the intial state $|C_4\rangle$ the expectation
value of the final state after dephaing with respect to the entanglement witness, 
$\mathcal{W}_{C4}$, is given by $-\frac{1}{4}\left(p^2-4p+2\right)$. Thus,  
cluster state entanglement can be detected by the entanglement witness 
for $p < 2-\sqrt{2} \simeq .586$. Interestingly, the expectation value with respect 
to the entanglement witness 
is equal to $N_{12}$, the most negative eigenvalue of the partial transpose of the 
final state with respect to qubits 1 and 2, which thus exhibits ESD at the same value. 
$N_1$, $N_{13}$, and $N_{14}$ do not undergo ESD. $N_1$, the lowest eigenvalue of 
the partial transpose of the final state with respect to one qubit, is given by 
$-\frac{1}{4}\left(p^2-3p+2\right)$. The most negative eigenvalues of the partial 
transpose of the state with respect to qubits 1 and 3 ($N_{13}$) and 1 and 4 ($N_{14}$) 
are four times degenerate and given by $\frac{1}{4}(p-1)$. 
Non-zero negativity for only some qubit partitions implies the presence of bound 
entanglement. For the initial state $|C_4\rangle$ under dephasing bound entanglement is 
present in the state for $p \agt .586$. 

For the intial state $|C_{4H}\rangle$ $N_1 = N_{12}$ with the most negative eignevalue of
the partial transpose of the final state given by $\frac{1}{16}\left(-4-4\tilde{p}^3+6p-p^2\right)$,
where $\tilde{p} = \sqrt{1-p}$. 
Both exhibit ESD at $p = -2+2\sqrt{2} \simeq .828$. For $N_{13}$ the most negative eigenvalue
is given by $\frac{1}{16}\left(-4\tilde{p}+2p-p^2\right)$ and is 
the last negativity to exhibit ESD, which occurs when $p \simeq .938$. For $N_{14}$ the lowest eigenvalue
is doubly degenerate and given by $\frac{1}{16}\left(-4+4p+p^2\right)$. ESD is exhibited at
$p = -2+2\sqrt{2} \simeq .828$ which is the same dephasing value at which $N_1$ exhibits ESD.
Again note the presence of bound entanglement for $.828 \leq p \leq .938$.  
The expectation value of the final state with respect to the entanglement witness, 
$\mathcal{W}_{C4H}$ is given by $\frac{1}{16}\left(-8\tilde{p}+p(8+4\tilde{p}-p)\right)$,
Thus, the witness fails to detect entanglement for 
$p > 2(-\sqrt{2}+2^{3/4}) \simeq .535$. The evolution of the above entanglement 
metrics as a function of $p$ are shown in Fig.~\ref{C4}. 

\section{Final State Fidelity}

\begin{figure}
\includegraphics[width=5cm]{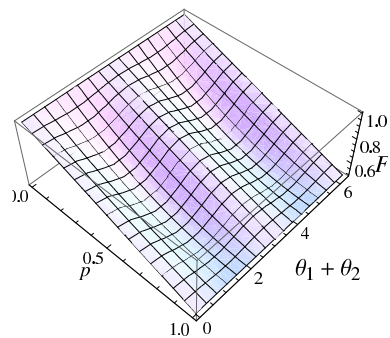}
\caption{(Color online) Fidelity of the state of the single 
unmeasured qubit from the four qubit cluster state $|C_4\rangle$ 
as a function of the dephasing strength $p$, and the sum of the first 
two measurement angles $\theta_1+\theta_2$. The third measurement 
angle $\theta_3$ does not affect the fidelity. The unmeasured qubit 
is the final state of the cluster computational logical qubit after 
performance of an arbitrary single qubit rotation via measurement. 
There is no sign of any sort of discontinuity that might have 
been expected due to ESD at $p \simeq .586$.}
\label{F4}
\end{figure}

Having observed that some sort of ESD occurs for both of our chosen representations 
of the four qubit cluster state, we now seek to determine whether ESD affects the 
utilization of the cluster state as a means of implementing a general single qubit 
rotation in the measurement based cluster model of quantum computation. To implement such 
a rotation measurements at an angle $\theta_t$ with respect to the positive $x$
axis in the $x-y$ plane are performed on the first three qubits, $t = 1,2,3$, giving
a one qubit final state as a function of the measurement angles and the dephasing strength, 
$\rho_f(p,\theta_1,\theta_2,\theta_3)$. We look at the fidelity of the state of the unmeasured 
qubit as compared to the same state without dephasing: 
\begin{equation}
F_r(p,\theta_1,\theta_2,\theta_3) = {\rm{Tr}}[\rho_f(p,\theta_1,\theta_2,\theta_3)\rho_f(0,\theta_1,\theta_2,\theta_3)].
\end{equation}
For convenience we have assumed that the outcome of each measurement is $-1$ in 
the chosen measurement basis, such that $m = 0$ and no extra $X$ rotations are necessary. 
A measurement of $+1$ would simply add the necessity for an $X$ rotation. We note that 
the fidelity calculation was done only for initial states $|C_4\rangle$ and 
$|C_{4H}\rangle$ while full process tomography is needed to completely 
determine the dynamics of the single qubit rotation.   

For initial state $|C_4\rangle$ the fidelity can be determined analytically,
\begin{eqnarray}
\label{eqFC4}
F_{C4}(p,\theta_1,\theta_2,\theta_3) &=& \frac{1}{4}(4+p(p-3) \nonumber\\
						 &+& p(1-p)\cos(2(\theta_1+\theta_2))).
\end{eqnarray}
Notice that for this representation, $\theta_3$ cancels and the other measurement angles 
contribute only as $\theta_1+\theta_2$. The fidelity is plotted in Fig.~\ref{F4} and
shows an oscillating plane steadily and smoothly decreasing toward, but never reaching, 
$F_{C4} = .5$. The amplitude 
of the oscillations decrease at high and low values of $p$ and reach a maximum 
at $p \simeq .5$. We do not see any sort of sharp transition or discontinuity in the 
behavior of $F_{C4}$ at $p \simeq .586$ as one might expect due to the sudden 
disappearance of $N_{12}$ for the complete four qubit cluster. 

As mentioned above, the initial state $|C_4\rangle$ undergoes ESD only with respect 
to $N_{12}$. One may suggest that the reason ESD is not manifest in the fidelity degradation 
of the unmeasured qubit for this initial state is because there is still some entanglement, 
$N_1$, which does not exhibit ESD, present in the state. 
To explore this we now look at the initial state $|C_{4H}\rangle$ which, 
under dephasing, exhibits ESD for all negativity measures. Following the above, we find the 
fidelity of the final single qubit state as a function 
of $p$ and measurement angles $\theta_t, t = 1,2,3$ for the intial state $|C_{4H}\rangle$ to be:
\begin{widetext}
\begin{eqnarray}
F_{C4H}(p,\theta_1,\theta_2,\theta_3) &=& \frac{1}{64}(2\tilde{p}^3\cos(2(\theta_1-\theta_2))
	+4p^{\prime}\cos(2\theta_2)+2\tilde{p}^3\cos(2(\theta_1+\theta_2))+2p^{\prime}\cos(2(\theta_1-\theta_3)) \nonumber\\
	&+& \tilde{p}^3\cos(2(\theta_1-\theta_2-\theta_3))+2p^{\prime}\cos(2(\theta_2-\theta_3))
	+\tilde{p}^3\cos(2(\theta_1+\theta_2-\theta_3)) \nonumber\\
	&+& 4(p-1)\cos(2\theta_1)\left(\tilde{p}-2(p+1)\cos\theta_3^2\right)
	+12p^{\prime}\cos(2\theta_3)+4(11+5\tilde{p}^3+3\cos(2\theta_3)) \nonumber\\
	&+& 2p^{\prime}\cos(2(\theta_1+\theta_3))+\tilde{p}^3\cos(2(\theta_1-\theta_2+\theta_3))
	+2p^{\prime}\cos(2(\theta_2+\theta_3))+\tilde{p}^3\cos(2(\theta_1+\theta_2+\theta_3)) \nonumber\\
	&+& 16\cos(2\theta_2)\cos\theta_3^2\sin\theta_1^2+8p^2(\cos\theta_3^2(1+2\cos(2\theta_2)\sin\theta_1^2)
	-\cos\theta_2\sin(2\theta_1)\sin(2\theta_3)) \nonumber\\
	&+& 8p(-4\cos\theta_3^2(1+\cos(2\theta_2)\sin\theta_1^2)+\cos\theta_2\sin(2\theta_1)\sin(2\theta_3))).
\end{eqnarray}
\end{widetext}

\begin{figure}[t]
\includegraphics[width=4cm]{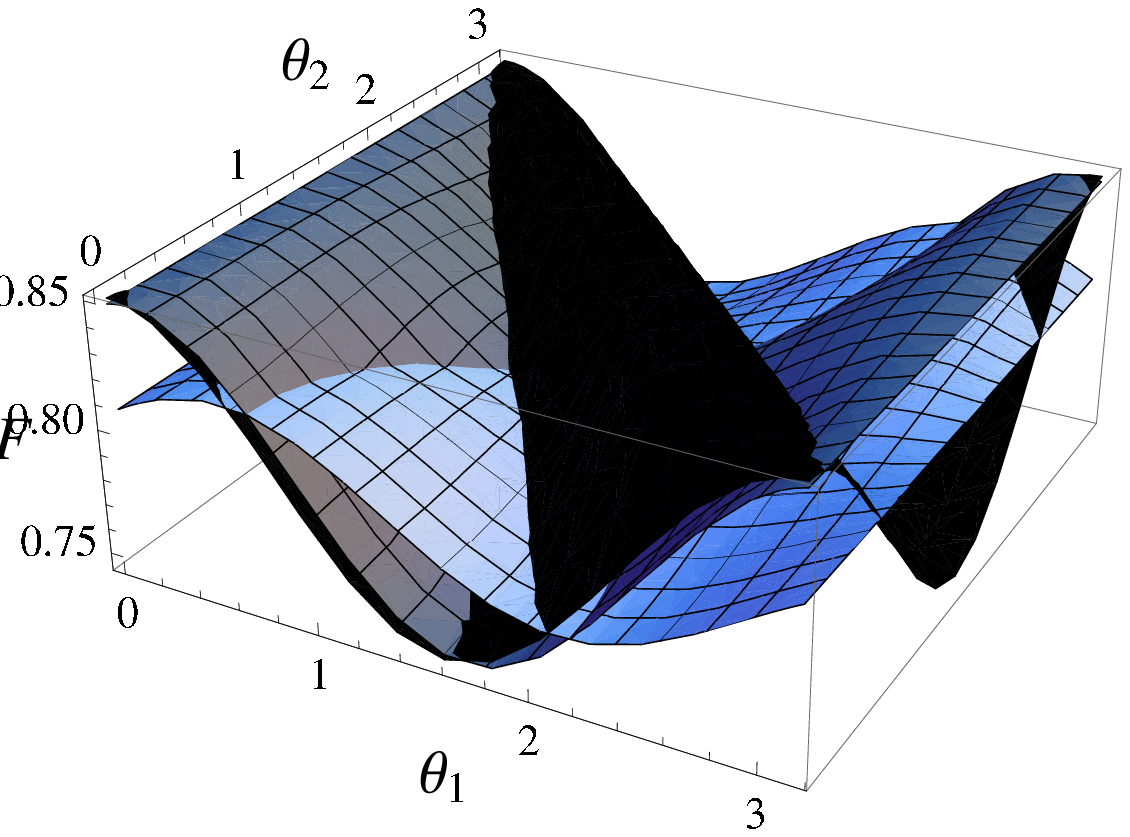}
\includegraphics[width=4cm]{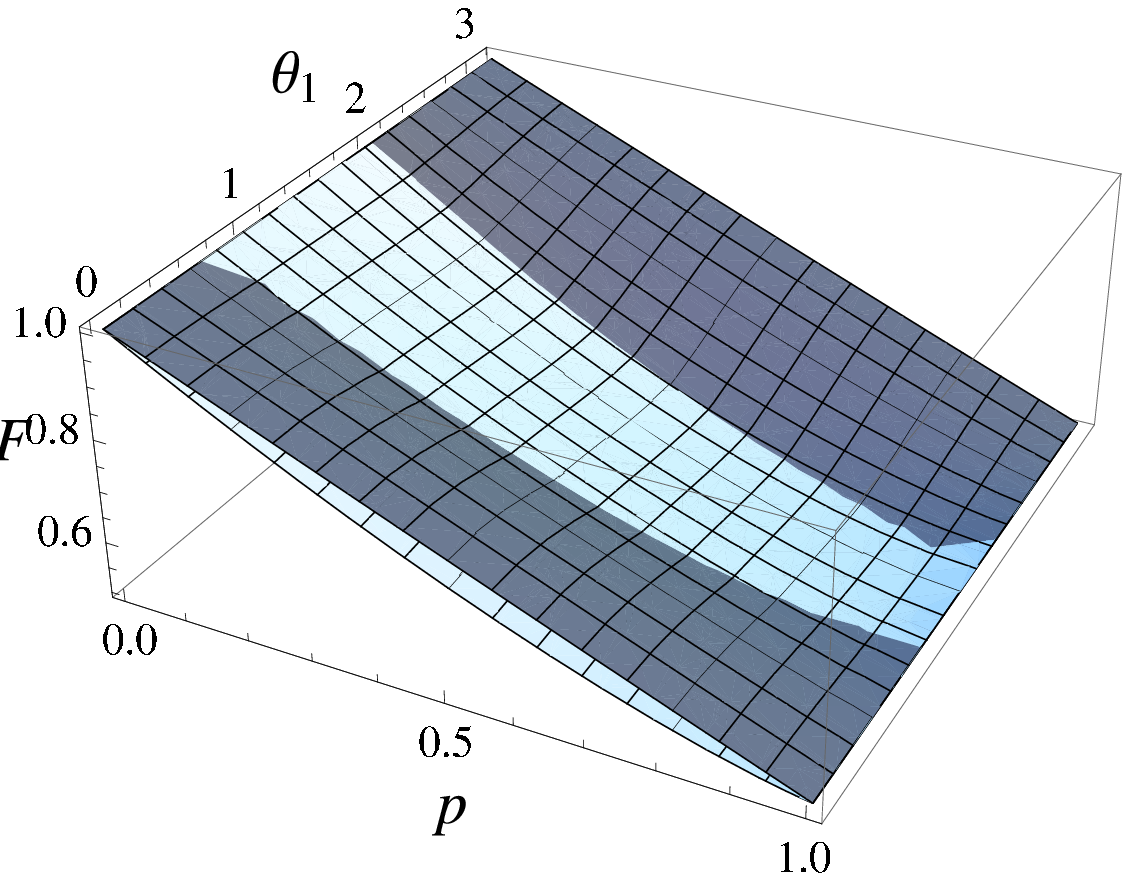}
\caption{(Color online) Left: using intial state $|C_{4H}\rangle$, fidelity of the 
state of the single unmeasured qubit such that an arbitrary rotation 
has been performed via the cluster 
state as a function of two of the measurement angles $\theta_1$ and 
$\theta_2$. The curves are $\theta_3 = \pi/16$ (gray) and 
$\pi/3$ (light) for $p = .3$. The black curve is the fidelity of the 
state of the single unmeasured qubit with dephasing 
for the intial state $|C_{4}\rangle$. This is plotted so as to compare the 
range of fidelities of the two initial states given the same evolution. 
Right: fidelity as a function of 
dephasing strength and $\theta_1$ with $\theta_2 = \pi/4$ and the two 
curves again equal to $\theta_3 = \pi/16$ (gray) and 
$\pi/3$ (light). As a function of $p$ we see the overall fidelity decreases
steadily toward .5 without any discontinuity.}
\label{F4H}
\end{figure}

Fig.~\ref{F4H} plots the fidelity as a function of the three measurement 
angles and $p$ (see figure caption). 
As a function of $p$ the fidelity decreases 
almost uniformly approaching, but not reaching, $F_{C4H} = .5$. Again we do not 
see any discontinuity or change of behavior at the dephasing strengths where 
ESD is exhibited for the complete cluster state, $p \simeq .828$ and 
$p \simeq .938$. 

Fig.~\ref{F4H} (left) also shows the fidelity of the 
state of the single unmeasured qubit for the intial state 
$|C_{4}\rangle$ and dephasing strength $p = .3$ as a function of 
the measurement angles. Note that the range of fidelity is
the same for both initial states but the maximum and minimum points
as a function of measurement angle are different. The equivalent fidelity 
range for the two cluster representations is in 
contrast to the disappearance of entanglement which occurs at
different dephasing strengths for the two cluster state representations.

\section{Two Qubit Fidelities and Concurrence}

So far our exploration of fidelity decay and entanglement as functions 
of dephasing indicate that ESD does not affect the utility of a cluster 
state as a means of implementing an arbitrary logical single qubit 
rotation. However, the picture changes when we explore fidelities and 
sudden bi-partite entanglement death of two qubits after having measured 
the other two qubits. To quantify the bi-partite entanglement between 
the two unmeasured qubits I use the concurrence
\cite{conc}, $C_{jk}$. The concurrence between two qubits $j$ and $k$ with 
density matrix $\rho_{jk}$ is usually defined as the maximum of zero and $\Lambda$, where 
$\Lambda = \sqrt{\lambda_1}-\sqrt{\lambda_2}-\sqrt{\lambda_3}-\sqrt{\lambda_4}$
and the $\lambda_i$ are the eigenvalues of 
$\rho_{jk}(\sigma_y^j\otimes\sigma_y^k)\rho_{jk}^*(\sigma_y^j\otimes\sigma_y^k)$
in decreasing order. $\sigma_y^i$ is the $y$ Pauli matrix of qubit $i$. For the 
purposes of clearly seeing at what point ESD occurs we will use $\Lambda$ as
the concurrence noting that ESD occurs when $\Lambda = 0$ in finite time 
(i.~e.~before $p \rightarrow 1$).

\begin{figure}
\includegraphics[width=5cm]{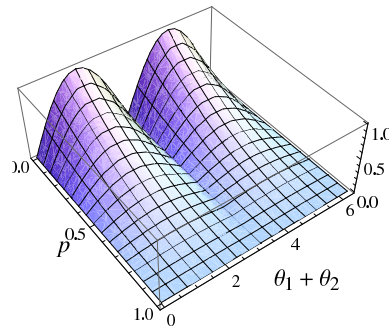}
\caption{(Color online) The concurrence between unmeasured 
qubits 3 and 4 after measurement on the first two qubits 
having started with the state $|C_4\rangle$. The concurrence
is plotted as a function of dephasing strength and measurement axes (which contribute
only as $\theta_1+\theta_2$). There is no ESD exhibited for this concurrence function.  
}
\label{C34}
\end{figure}

We start with the intial state $|C_4\rangle$, with measurements performed 
on qubits 1 (along the axis $\theta_1$) and 2 (along the axis $\theta_2$). 
The fidelity of the state of the two remaining 
qubits as a function of dephasing is given by Eq.~(\ref{eqFC4}), the fidelity of 
the final state of the fourth qubit after measurement on qubits 1, 2, and 3. This is so because 
eq.~(\ref{eqFC4}) does not depend on $\theta_3$. The concurrence between unmeasured 
qubits 3 and 4 is a function only of the sum of
the two measurement angles, $\theta_1 + \theta_2$, and $p$, and is given by 
$C^{C4}_{34} = \frac{1}{2\sqrt{2}}(\sqrt{A_{34}+B_{34}}-\sqrt{A_{34}-B_{34}})$ where:
\begin{eqnarray}
A_{34} &=& 2+p(p-2)(p-1)^2 \\
	 &-& (p-1)^2(2+p(p-2)\cos(2(\theta_1+\theta_2)), \nonumber
\end{eqnarray}
and
\begin{eqnarray}
B_{34} &=& 2(-2(p-1)^4(-1+p(p-2) \\
	 &+& (p-1)^2\cos(2(\theta_1+\theta_2)))\sin(\theta_1+\theta_2)^2)^{1/2}. \nonumber
\end{eqnarray}
The concurrence is plotted in Fig. \ref{C34}. We note that the fidelity of the state of the 
two unmeasured qubits never falls below .5 and no ESD is exhibited due to the dephasing.

\begin{figure}
\includegraphics[width=5cm]{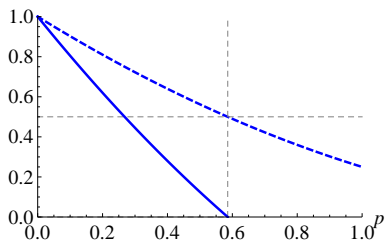}
\caption{\label{C4F24} (Color online)
Fidelity (dashed line) of the state of qubits 2 and 4 after measurements 
on qubits 1 and 3 of the initial state $|C_4\rangle$
compared to the concurrence (solid line) between these same qubits. Note that the 
fidelity crosses .5 (horizontal light line) at $p \simeq .586$ (vertical light line)
which is the same dephasing strength where ESD is exhibited by the concurrence between 
these two qubits and by $N_{12}$ of $|C_4\rangle$.
}
\end{figure}

If measurements are carried out on qubits 1 and 3 the fidelity of the state that 
remains on qubits 2 and 4 with dephasing is completely independent of 
any measurement angle and is given by $\frac{1}{4}(p-2)^2$. The concurrence
between qubits 2 and 4 after the measurements is also independent of measurement angle
and is given by $\frac{1}{2}(p^2-4p+2)$. Note that the fidelity goes to .5 and the 
concurrence goes to zero at $p = 2 - \sqrt{2} \simeq .586$, the same value for which 
$N_{12}$ of the four qubit cluster state exhibits ESD and the expectation value of 
the four qubit state with respect to $\mathcal{W}_{C4}$ goes to zero. While there is no
discontinuity in the fidelity behavior at the dephasing strength that causes ESD, 
the fidelity does cross the critical value of .5 at the same dephasing strength. Thus,
ESD indicates the severity of the decreased correlation between the dephased and not 
dephased state. The correlation between these metrics is shown in Fig.~\ref{C4F24}.
Also note that in the previous case, where qubits 1 and 2 are measured, there is no 
exhibition of ESD and the fidelity never reaches .5. Measurement on qubits 1 and 4 or 
qubits 2 and 3 give the exact same results as the measurements on 1 and 3.

\begin{figure}
\includegraphics[width=4cm]{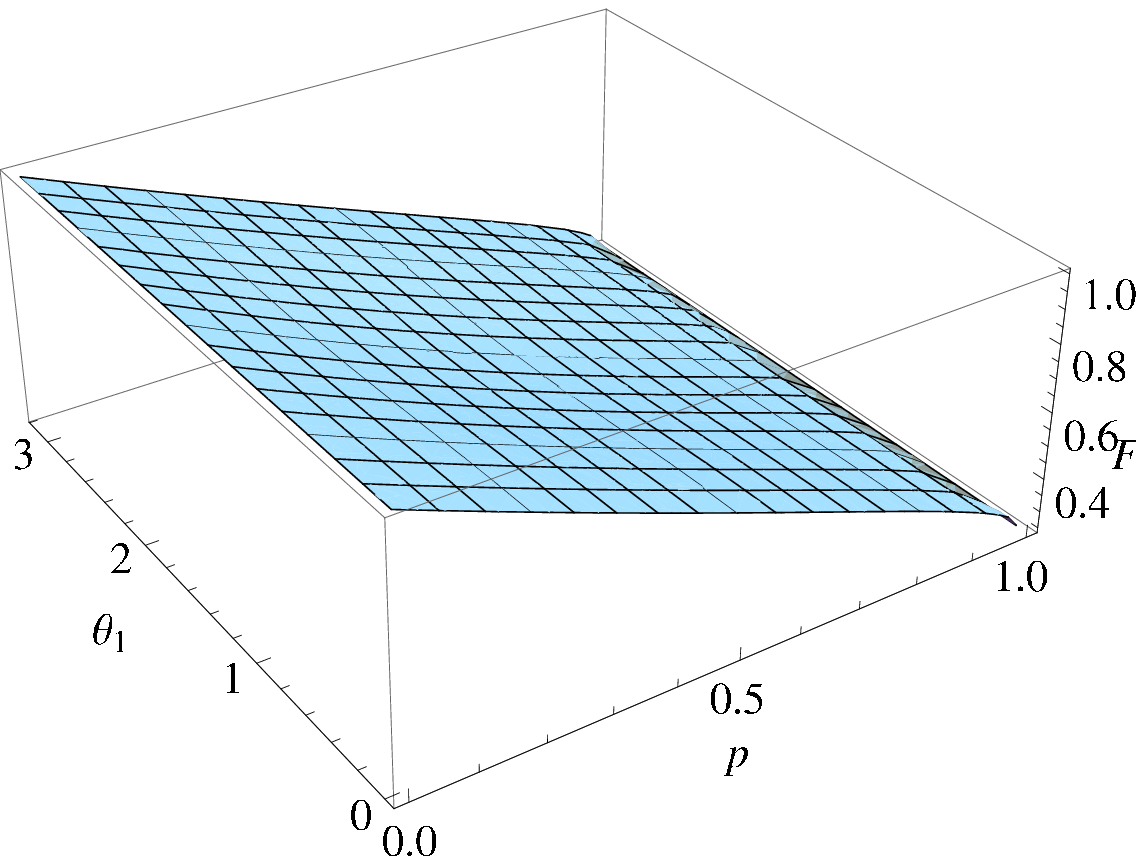}
\includegraphics[width=4cm]{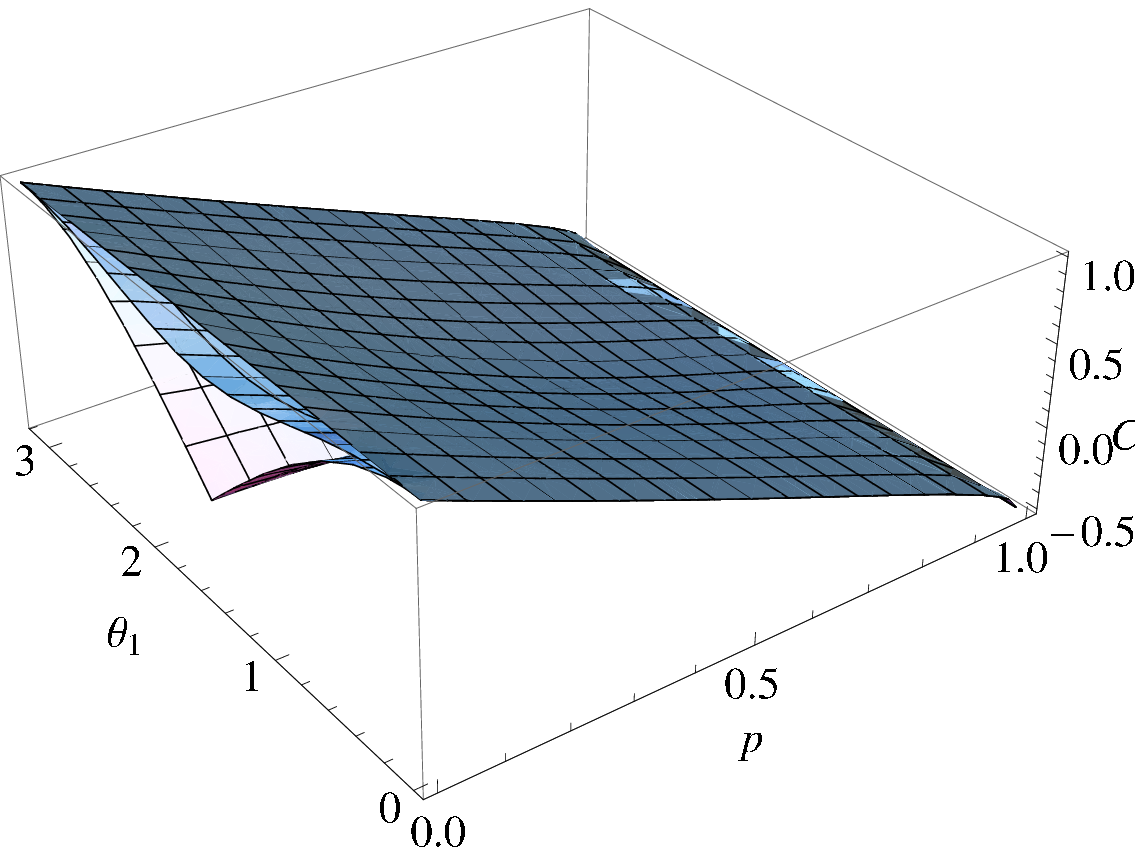}
\includegraphics[width=4cm]{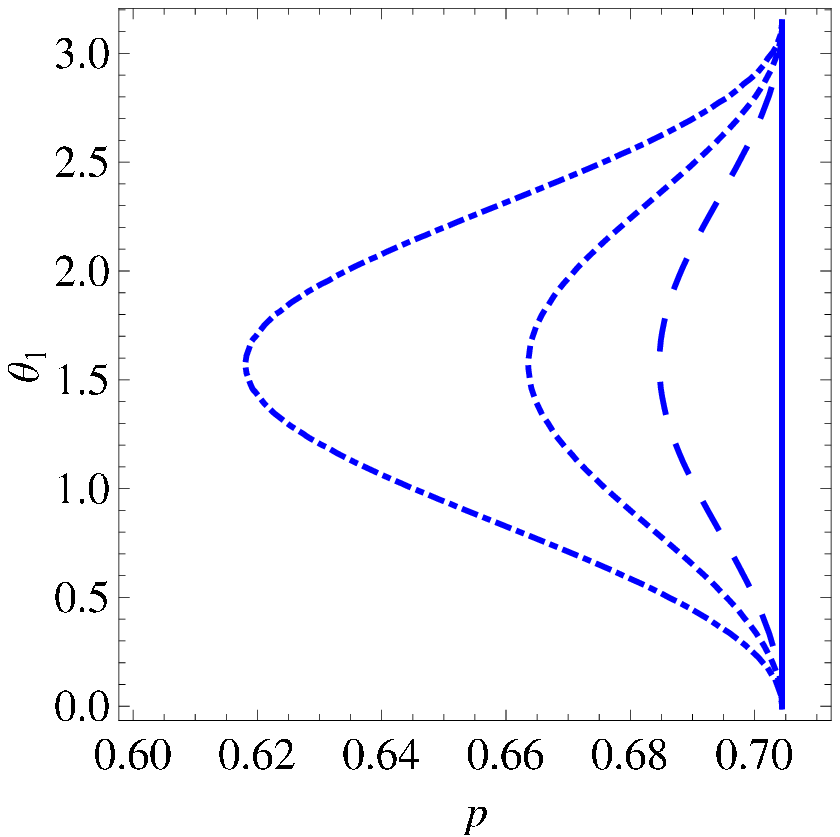}
\includegraphics[width=4cm]{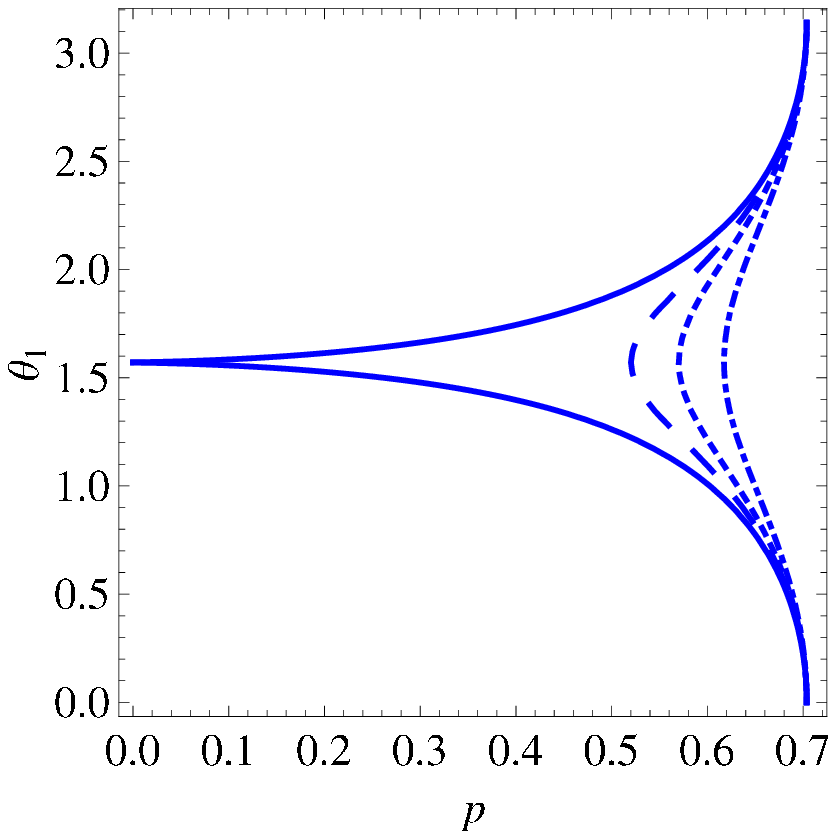}
\caption{\label{C4H34} (Color online)
Fidelity (left) of the state of qubits 3 and 4 after measurement on qubits 1
and 2 and concurrence (right) between those qubits
as a function of dephasing strength $p$ and the measurement axes angles. 
Top left: fidelity as a function of $p$ and $\theta_1$ for $\theta_2 = \pi/4$. 
Bottom left: contours of fidelity equal to .5 for $\theta_2 = 0$ (chained line), 
$\theta_2 = \pi/4$ (dotted line), $\theta_2 = \pi/3$ (dashed line), and $\theta_2 = \pi/2$
(solid line). The fidelity in all cases converges to $p \simeq .704$ as $\theta_1$ goes
to zero.
Top right: concurrence as a function of $p$ and $\theta_1$ for $\theta_2 = 0$
(bottom), $\theta_2 = \pi/4$ (middle), and $\theta_2 = \pi/2$ (top). 
Bottom right: contours of concurrence equal to zero showing where ESD occurs
(values of $\theta_2$ as in previous contour plot). 
The dephasing values at which ESD is exhibited approach .704, the exact value 
for which the fidelity goes to .5.
}
\end{figure}

We see similar correlations between fidelity and entanglement metrics when measuring 
cetain pairs of qubits of the inital state $|C_{4H}\rangle$. The fidelity of the  
state of qubits 3 and 4 upon measuring qubits 1 and 2 is given by:
\begin{eqnarray}
F_{34}(p,\theta_1,\theta_2) &=& \frac{1}{16}(8(1+\tilde{p})+p(-5-4\tilde{p}+p) \\
				    &-& p(p-1)(\cos(2\theta_1)-2\cos(2\theta_2)\sin\theta_1^2)). \nonumber
\end{eqnarray}
As shown in Fig.~\ref{C4H34}, when $p \simeq .704$ the fidelity goes to .5 as $\theta_1$ approaches 0 or $\pi$ 
or when $\theta_2$ approaches $\frac{\pi}{2}$. This is also the maximum dephasing value for
which we find ESD of the concurrence between unmeasured qubits 3 and 4 as shown in the figure
(we do not have an analytical solution for the concurrence). Thus, while once again we do not 
have a change of fidelity behavior due to ESD, the sudden death of concurrence does 
indicate the lowering of fidelity to the critical value of .5.  

\begin{figure}
\includegraphics[width=4cm]{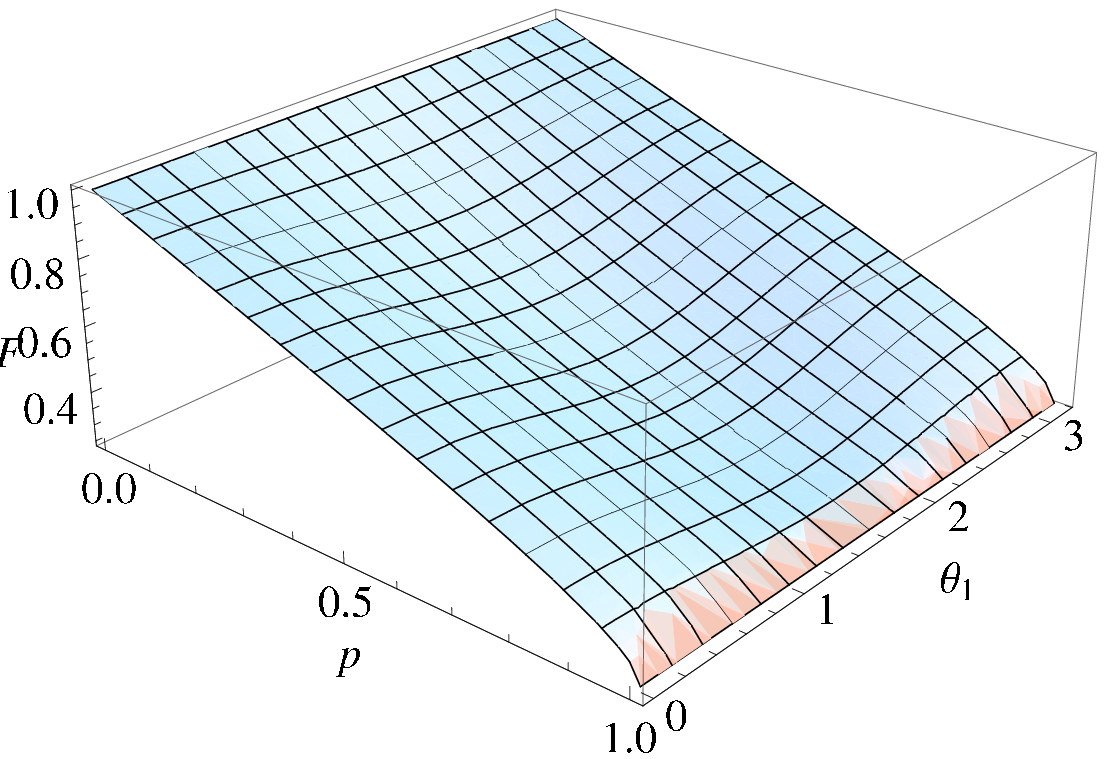}
\includegraphics[width=4cm]{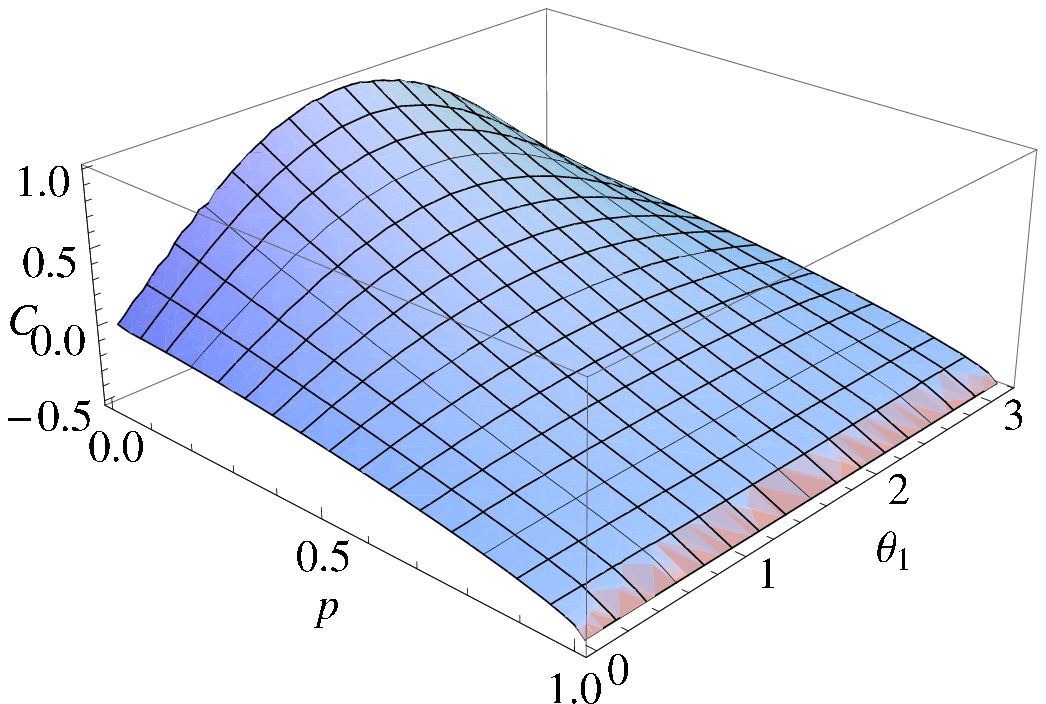}
\includegraphics[width=4cm]{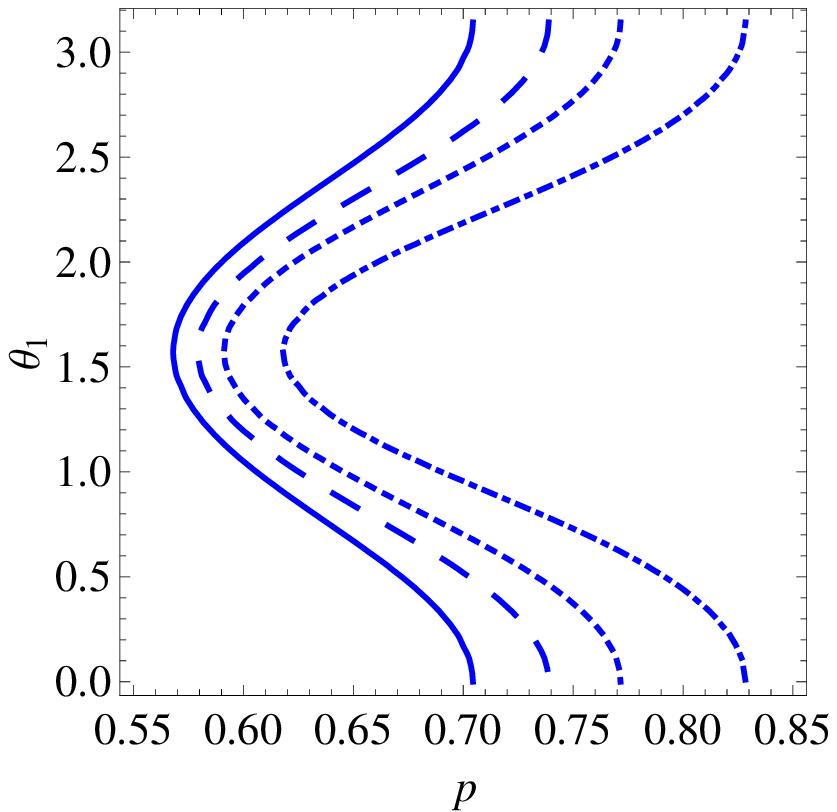}
\includegraphics[width=4cm]{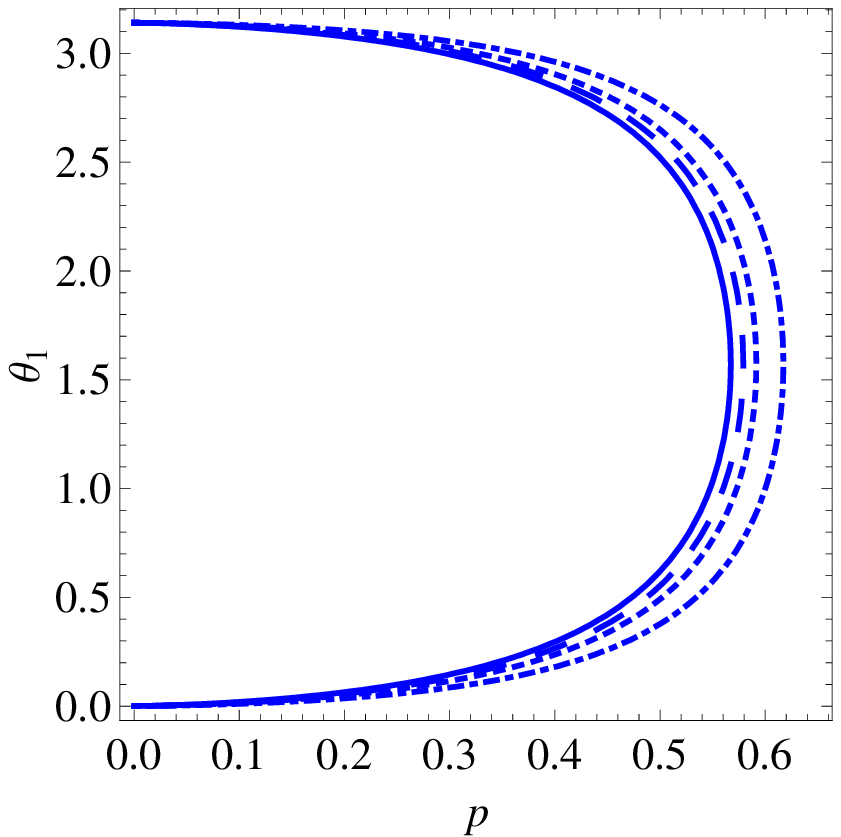}
\caption{\label{C4H24} (Color online)
Fidelity (left) of the state of qubits 2 and 4 after measurement on qubits 1
and 3 and concurrence (right) between those qubits
as a function of dephasing strength $p$ and measurement axes angles. 
Top left: fidelity as a function of $p$ and $\theta_1$ for $\theta_3 = \pi/4$. 
Bottom left: contours of fidelity equal to .5 for $\theta_3 = 0$ (chained line), 
$\theta_3 = \pi/4$ (dotted line), $\theta_3 = \pi/3$ (dashed line), and $\theta_3 = \pi/2$
(solid line). As $\theta_1$ and $\theta_3$ go to zero the fidelity equals .5 contour
goes to $p \simeq .828$, the value at which the state $|C_{4H}\rangle$ 
exhibits ESD for a number of entanglement measures.
Top right: concurrence as a function of $p$ and $\theta_1$ for $\theta_3 = \pi/4$. 
Bottom right: contours of concurrence equal to zero showing where ESD occurs
(values of $\theta_3$ as in previous contour plot). The maximum dephasing value at which 
ESD is exhibited is .618.
}
\end{figure}

The fidelity of the state of qubits 2 and 4 upon measuring qubits 1 and 3 
is given by:
\begin{eqnarray}
F_{24}(p,\theta_1,\theta_3) &=& \frac{1}{16}(8(1+\tilde{p})+p(-5-4\tilde{p}+p) \\
				    &+& p\cos(2\theta_1)(1+2\tilde{p}-p) \nonumber\\
				    &+& 2p\cos(2\theta_3)(\tilde{p}+(p-1)\sin\theta_1^2))\nonumber
\end{eqnarray}
and is plotted in Fig.~\ref{C4H24} along with the concurrence between unmeasured qubits 2 and 4. 
There does not appear to be a correlation between the fidelity and concurrence with respect to these
two unmeasured qubits. However, the maximum $p$ at which the fidelity crosses .5, 
when $\theta_1 = \theta_3 = 0$, is $2\sqrt{2} - 2 \simeq .828$, the exact value 
where the four qubit state $|C_{4H}\rangle$ exhibits ESD for $N_1$, $N_{13}$, and $N_{14}$. The 
minimum value at which the fidelity crosses .5 is at .568. The maximum $p$ at which ESD of 
concurrence is exhibited is .618.

\begin{figure}
\includegraphics[width=4cm]{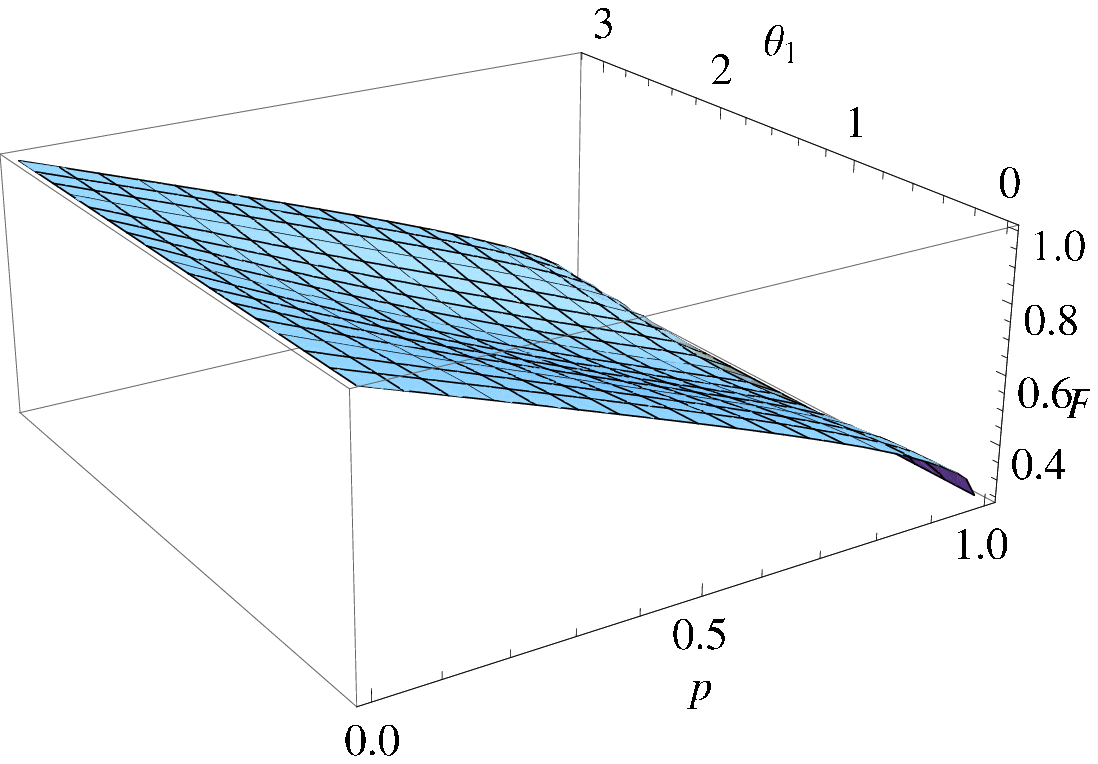}
\includegraphics[width=4cm]{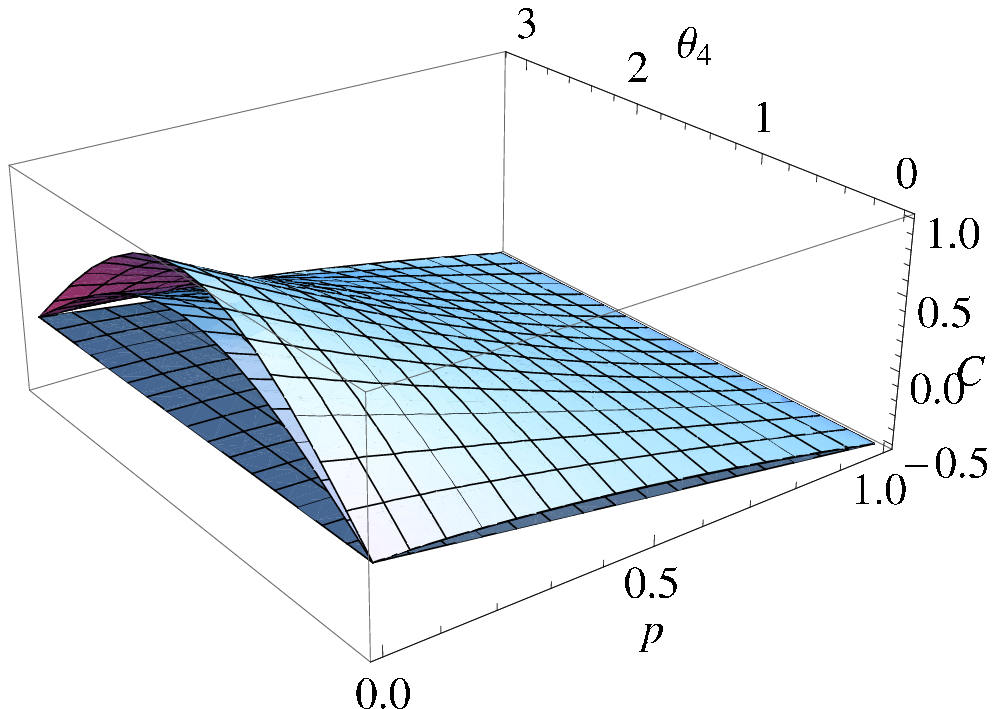}
\includegraphics[width=4cm]{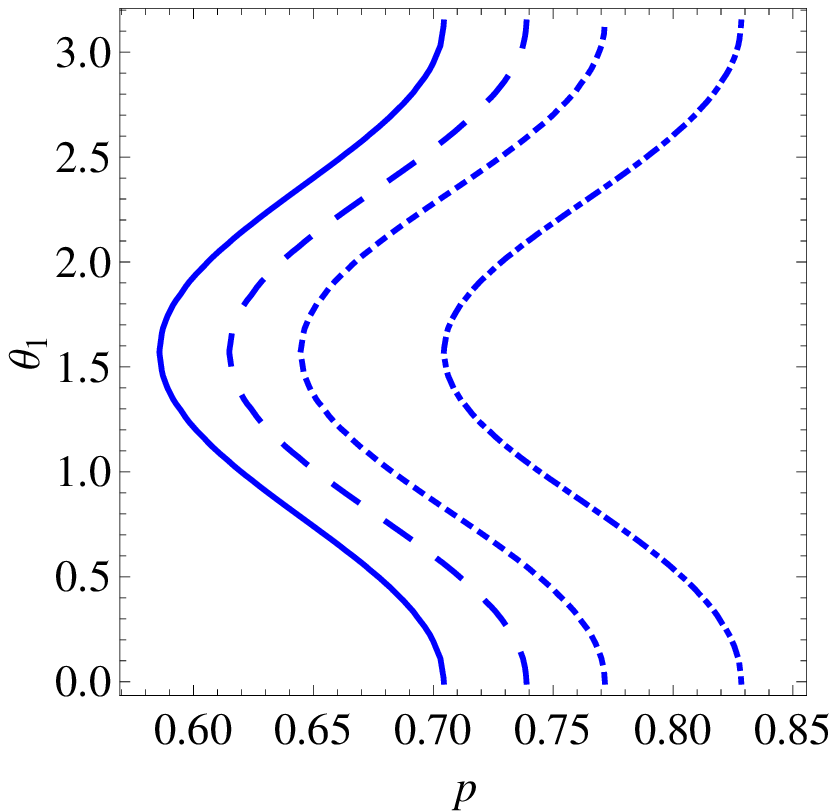}
\includegraphics[width=4cm]{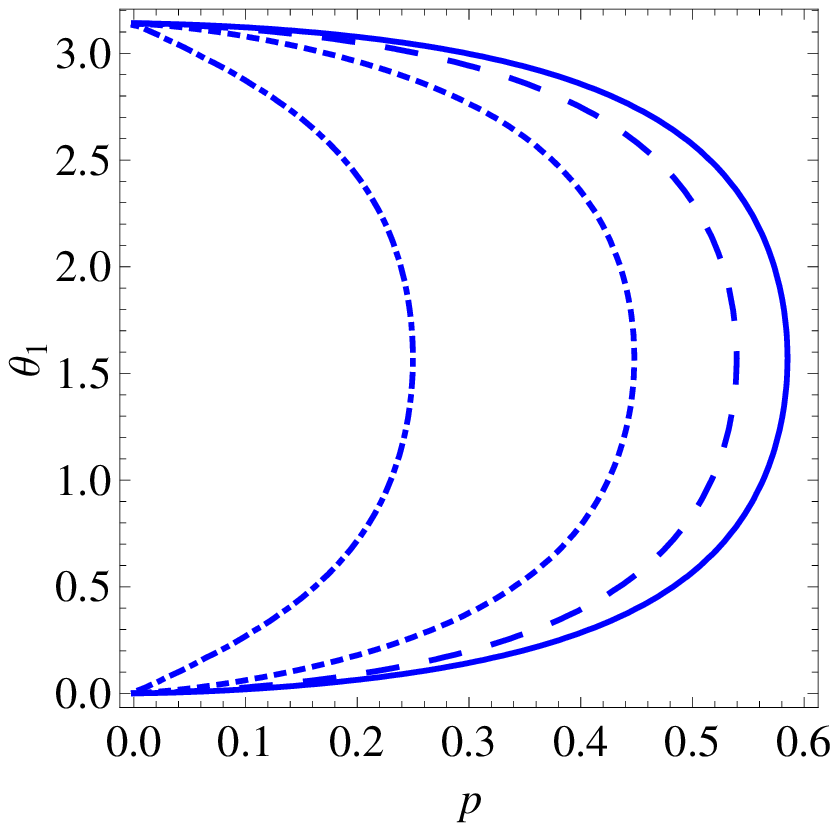}
\caption{\label{C4H23} (Color online)
Fidelity (left) of the state of qubits 2 and 3 after measurement on qubits 1
and 4 and concurrence (right) between those qubits
as a function of dephasing strength $p$ and measurement axes angles. 
Top left: fidelity as a function of $p$ and $\theta_1$ for $\theta_4 = \pi/4$. 
Bottom left: contours of fidelity equal to .5 for $\theta_4 = 0$ (chained line), 
$\theta_4 = \pi/4$ (dotted line), $\theta_4 = \pi/3$ (dashed line), and $\theta_4 = \pi/2$
(solid line). As $\theta_1$ and $\theta_4$ go to zero the fidelity equals .5 contour
goes to $p \simeq .828$, the value at which the state $|C_{4H}\rangle$ 
exhibits ESD for a number of entanglement measures.
Top right: concurrence as a function of $p$ and $\theta_4$ for $\theta_1 = 0$ (bottom),
$\theta_1 = \pi/4$ (middle), and $\theta_1 = \pi/2$ (top).  
Bottom right: contours of concurrence equal to zero showing where ESD occurs
for $\theta_4 = \pi/32$ (chained line), $\theta_4 = \pi/8$ (dotted line), 
$\theta_4 = \pi/4$ (dashed line), and $\theta_4 = \pi/2$ (solid line). 
The maximum dephasing value at which ESD is exhibited is .586 which
is the value at which ESD is exhibited for the state $|C_4\rangle$.
}
\end{figure}

The fidelity of the state of qubits 2 and 3 upon measuring qubits 1 and 4 is given by:
\begin{eqnarray}
F_{23}(p,\theta_1,\theta_4) &=& \frac{1}{16}(10+6\tilde{p}-p(7+2\tilde{p}-p)+\cos(2\theta_4) \nonumber\\
				    &\times& (-2+2\tilde{p}+3p-p^2)+ 2\cos(2\theta_1) \\
				    &\times& (\tilde{p}-\tilde{p}^3\cos(2\theta_4)-(p-2)(p-1)\sin\theta_4^2)),\nonumber
\end{eqnarray}
and plotted in Fig.~\ref{C4H23} along with the concurrence between unmeasured qubits 2 and 3. 
As in the previous case ESD may be an indicator of fidelity. The maximum $p$ at which 
the fidelity crosses .5, which occurs for 
$\theta_1 = \theta_3 = 0$, is $2\sqrt{2} - 2 \simeq .828$, the exact value where the 
four qubit state $|C_{4H}\rangle$ exhibits ESD for a number of entanglement measures. 
The minimum $p$ at which the fidelity crosses .5 is $2 - \sqrt{2} \simeq .586$, 
which is also equal to the the maximum $p$ at which ESD of concurrence is exhibited.
Though the initial state in this example was $|C_{4H}\rangle$ this is the value at 
which ESD occurs for the initial state $|C_4\rangle$. Such cross-correlation between 
the different cluster state representations can come from the measurements: measuring
some of the qubits at certain angles transforms the state from one represenation to 
the other.

\begin{figure}
\includegraphics[width=4cm]{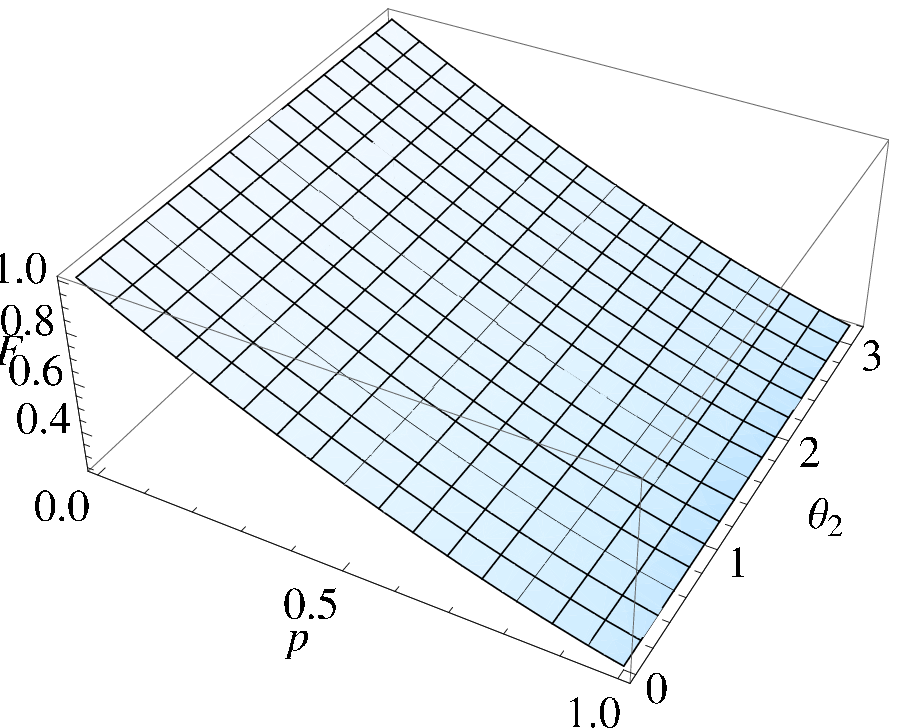}
\includegraphics[width=4cm]{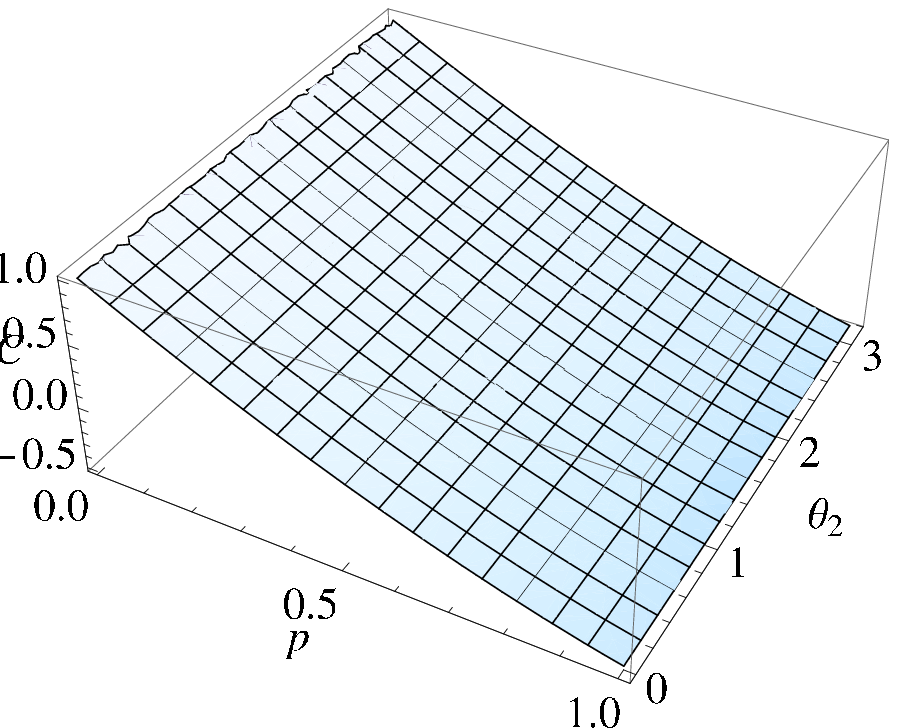}
\includegraphics[width=4cm]{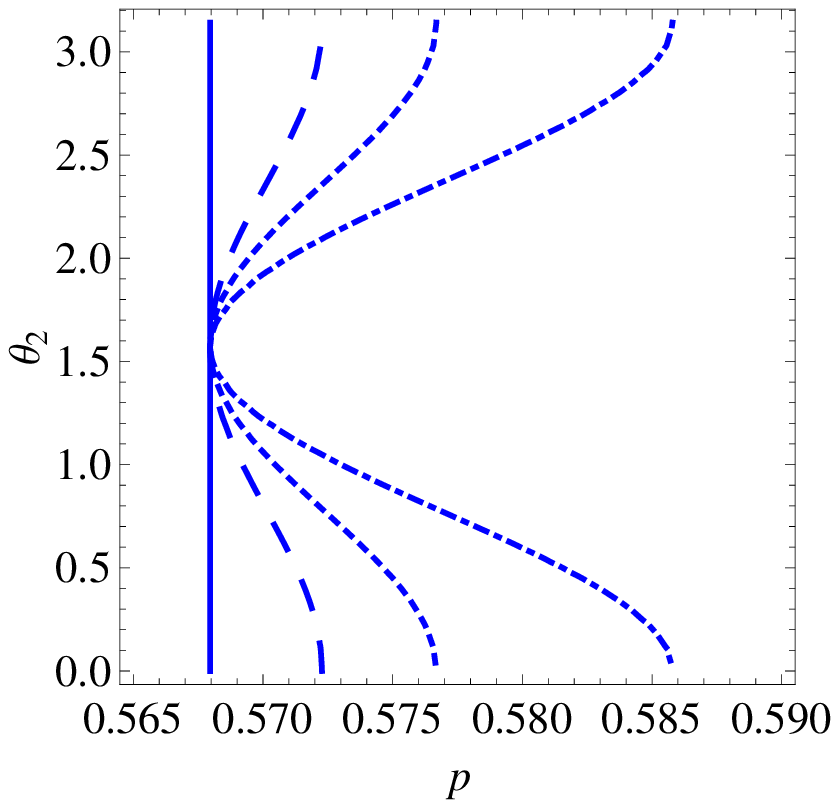}
\includegraphics[width=4cm]{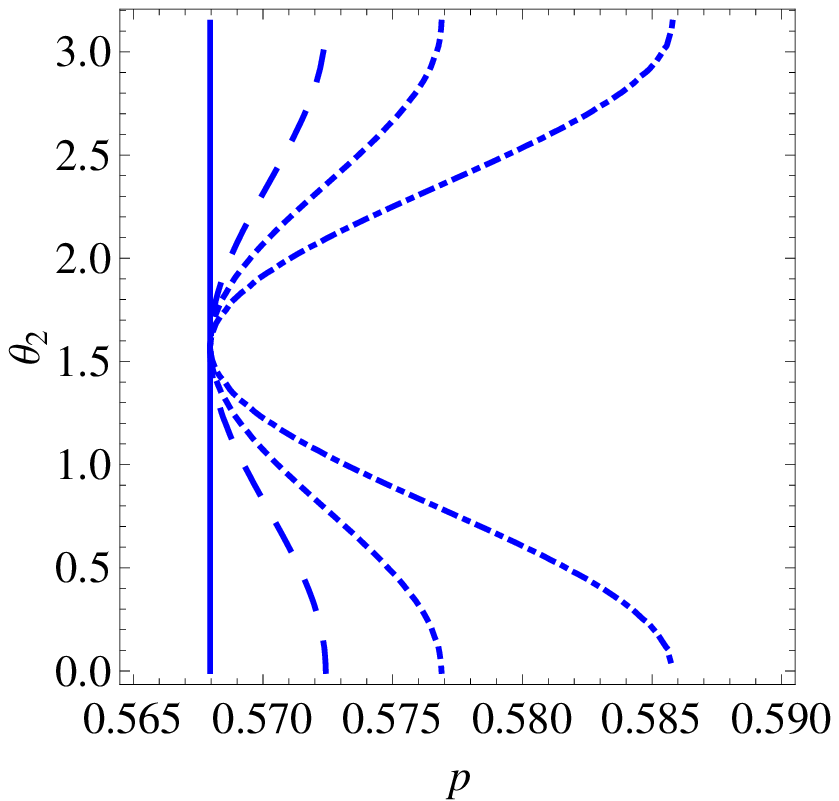}
\caption{\label{C4H14} (Color online)
Fidelity (left) of the state of qubits 1 and 4 after measurement on qubits 2
and 3 and concurrence (right) between those qubits
as a function of dephasing strength $p$ and the measurement axes angles. 
Top left: fidelity as a function of $p$ and $\theta_2$ for $\theta_3 = \pi/4$. 
Bottom left: contours of fidelity equal to .5 for $\theta_3 = 0$ (chained line), 
$\theta_3 = \pi/4$ (dotted line), $\theta_3 = \pi/3$ (dashed line), and $\theta_3 = \pi/2$
(solid line). As $\theta_3$ goes to $\pi/2$ the fidelity equals .5 contour
goes to $p \simeq .568$, when $\theta_3$ and $\theta_2$ go to zero 
the fidelity equals .5 contour goes to $p \simeq .586$, the value at which the state $|C_{4H}\rangle$ 
exhibits ESD for a number of entanglement measures.
Top right: concurrence as a function of $p$ and $\theta_2$ for $\theta_1 = \pi/4$.  
Bottom right: contours of concurrence equal to zero showing where ESD occurs
(same $\theta_3$ values as above). These curves are equivalent to those of the 
fidelity equals .5 curves.
}
\end{figure}

The fidelity of the state of qubits 1 and 4 upon measuring qubits 2 and 3 
is given by:
\begin{eqnarray}
F_{14}(p,\theta_2,\theta_3) &=& \frac{1}{16}(10+6\tilde{p}-p(7+6\tilde{p}-p)\\
				    &+& (p-1)(-2+2\tilde{p}+p)(\cos(2\theta_2)\nonumber\\
				    &+& 2\cos\theta_2^2\cos(2\theta_3)).  \nonumber
\end{eqnarray}
The concurrence between unmeasured qubits 1 and 
4 is given by $\rm{Max}[-\frac{p}{2}-c_{14},-\frac{p}{2}+c_{14}]$ where 
\begin{eqnarray}
c_{14} &=& \frac{1}{4}((p-1)^2(16+p(p-16) \\
	 &+& p^2(\cos(2\theta_2)+2\cos\theta_2^2\cos(2\theta_3)))^{1/2}. \nonumber
\end{eqnarray}
Fig.~\ref{C4H14} demonstrates the strong correlation between the dephasing value 
where ESD is exhibited and the value where the fidelity goes to .5. Furthermore, the highest 
dephasing possible where ESD occurs (and when the fidelity goes to .5) is at 
$p = 2-\sqrt{2} \simeq .586$, the same value for which we find ESD for the state $|C_4\rangle$.
Again this points to the possibility of the measurement `transforming' between the two 
representations of the cluster state. The lowest dephasing at which ESD occurs 
(or where the fidelity goes to .5) is at .568.
All the above ESD and fidelity results are summarized in Table \ref{table}.


\begin{table*}
\caption{The three parts of the table show 1) the values of $p$ at which ESD occurs in four 
qubit cluster states subject to dephasing as measured by the expectation value of the proper 
entanglement witness ($\mathcal{W}_r$, $r = 4, 4H$), the negativity with partial transpose taken 
with respect to one ($N_1$) and two ($N_{1k}$, $k = 2,3,4$) qubits, 2) the value of $p$ for 
which the concurrence, $C_{jk}$, between two unmeasured qubits $j$ and $k$ of the four qubit cluster 
state goes to zero after measurments on the other two qubits, 3) the value of $p$ for which the 
fidelity of the dephased states goes to .5 for two qubit states $F_{jk}$, 
after measurement on the other two qubits, or the fidelity of the one qubit states $F_r$, 
after measurement on three qubits.
}
\begin{tabular}{||c||c|c|c|c|c||}
\hline 
 & $\mathcal{W}$ & $N_{1}$ & $N_{12}$ & $N_{13}$ & $N_{14}$ \\\hline
\hline
$|C_{4}\rangle$  & .586 & none & .586 & none & none \\\hline
$|C_{4H}\rangle$ & .535 & .828 & .828 & .938 & .828 \\\hline
\hline
& $C_{34}$ & $C_{24}$ & $C_{23}$ & $C_{14}$ & \\\hline
$|C_{4}\rangle$  & none & .586 & .586 & .586 & \\\hline
$|C_{4H}\rangle$ & $\leq .704$ & $\leq .618$ & $\leq .586$ & $.568\leq p \leq .586$ & \\\hline
\hline
& $F_{34}$ & $F_{24}$ & $F_{23}$ & $F_{14}$ & $F_r$ \\\hline
$|C_{4}\rangle$  & none & .586 & .586 & .586 & none \\\hline
$|C_{4H}\rangle$ & $.618\leq p \leq .704$ & $.568\leq p \leq .828$ & $.586\leq p \leq .828$ & $.568\leq p \leq .586$ & none \\\hline

\end{tabular}
\label{table}
\end{table*}

Finally, we note that in all of the above we have made a number of assumptions. First, 
we have assumed that the intial cluster state
is constructed perfectly. One way to relax this assumption is by looking at intial states
of the form:
$\rho_r = \frac{1-q}{16}\openone+q|C_r\rangle\langle C_r|$, where $r = 4,4H$. Preliminary 
explorations using this starting state indicate that there is merely a shift in entanglement values 
downwards but that there is no fundamental change in the behavior of the entanglement. Another assumption 
is that the dephasing strength is equal on all four qubits. This is unrealistic for a number
of reasons but especially so if not all of the measurements are performed at the same time 
(non-simultaneous measurements are necessary when trying to implement a given logical rotation 
because the measurement axes for a given qubit depends on the outcome of the measurement on the 
previous qubit). A way to relax this assumption without
significantly increasing the number of variables in the problem may be to add a $k\Delta p$ term 
to the dephasing strength where $\Delta p$ represents the dephaing the occurs during the time 
between subsequent measurements and $k$ is an integer.

\section{Conclusions}

In conclusion, I have studied the entanglement evolution of a four
qubit (chain) cluster state in a dephasing environment. Specifically,
I have looked at two represenations of the state differing by single
qubit rotations. Both of these representations exhibit entanglement
sudden death under sufficient dephasing. The difference in the dephasing
strength at which this occurs may be important when deciding in what 
representation to store a cluster state. The issue of storage is 
especially relevant during the construction of optical cluster states 
but may have relevance to other implementations as well. 

I asked whether ESD affects 
the utility of the cluster state in implementing a general single qubit
rotation in the cluster state measurement based quantum computation paradigm.
Judging from the fidelity decay of the single unmeasured qubit as a function 
of dephasing strength and the measurement axes angles of the three measurements 
the answer would seem to be no. I see no indication in the fidelilty behavior
that ESD has taken place. Instead the fidelity decreases smoothly with 
increased dephasing with no discontinuities or dramatic changes in 
behavior. However, there are clear correlations (sometimes total and sometimes
at certain limits) between the fidelity of the state of two qubits remaining from
the four qubit cluster state
after measurement on the other two qubits, and ESD of the negativity for the entire
cluster state or ESD of the concurrence between the said two unmeasured qubits. This
correlation does not appear as a discontinuity in the fidelity decay behavior but 
instead is manifest by the fidelity crossing the critical value of .5. 
Thus, we could say that ESD may be an {\em indicator} of how badly a certain cluster 
state operation was carried out. However, this is not the same as saying that ESD
itself negatively affects quantum information protocols. The question of whether ESD 
affects quantum information protocols requires further study and may be
related to the more general issue of the role of entanglement in quantum computation.

I thank G. Gilbert for helpful feedback and acknowledge 
support from the MITRE Technology Program under MTP grant \#07MSR205.

\end{document}